\documentstyle[prd,aps,preprint,psfig]{revtex}

\newcommand{\CL}{{\cal L}}

\newcommand{\CO}{{\cal O}}

\newcommand{\bear}{\begin{array}}  \newcommand{\eear}{\end{array}}
\newcommand{\bea}{\begin{eqnarray}}  \newcommand{\eea}{\end{eqnarray}}
\newcommand{\beq}{\begin{equation}}  \newcommand{\eeq}{\end{equation}}
\newcommand{\bef}{\begin{figure}}  \newcommand{\eef}{\end{figure}}
\newcommand{\bec}{\begin{center}}  \newcommand{\eec}{\end{center}}
\newcommand{\non}{\nonumber}  
\newcommand{\lmk}{\left(}  \newcommand{\rmk}{\right)}
\newcommand{\lkk}{\left[}  \newcommand{\rkk}{\right]}
\newcommand{\lhk}{\left \{ }  \newcommand{\rhk}{\right \} }
\newcommand{\del}{\partial}  

\newcommand{\bib}{\bibitem} 
\newcommand{\la}{\left\langle} \newcommand{\ra}{\right\rangle}

\def\IB#1#2#3{{\bf #1}, #2 (19#3)}
\def\IBB#1#2#3{{\bf #1}, #2 (20#3)}
\def\IBID#1#2#3{{\it ibid}. {\bf #1}, #2 (19#3)}
\def\IBIDD#1#2#3{{\it ibid}. {\bf #1}, #2 (20#3)}

\def\APJL#1#2#3{Astrophys. J. Lett. {\bf #1}, L#2 (19#3)}
\def\APJLL#1#2#3{Astrophys. J. Lett. {\bf #1}, L#2 (20#3)}

\def\NATT#1#2#3{Nature (London) {\bf #1}, #2 (20#3)}
\def\NPB#1#2#3{Nucl. Phys. {\bf B#1}, #2 (19#3)}

\def\PLB#1#2#3{Phys. Lett. B {\bf #1}, #2 (19#3)}

\def\PLBold#1#2#3{Phys. Lett. {\bf#1B}, #2 (19#3)}

\def\PRD#1#2#3{Phys. Rev. D {\bf #1}, #2 (19#3)}
\def\PRDD#1#2#3{Phys. Rev. D {\bf #1}, #2 (20#3)}

\def\PRL#1#2#3{Phys. Rev. Lett. {\bf#1}, #2 (19#3)}
\def\PRLL#1#2#3{Phys. Rev. Lett. {\bf#1}, #2 (20#3)}

\def\PTP#1#2#3{Prog. Theor. Phys. {\bf #1}, #2 (19#3)}

\begin{document}

\tighten
\draft
\title{Supersymmetric topological inflation model}
\author{M. Kawasaki}
\address{Research Center for the Early Universe, University of Tokyo,
  Tokyo 113-0033, Japan}
\author{Masahide Yamaguchi}
\address{Research Center for the Early Universe, University of Tokyo,
  Tokyo 113-0033, Japan}

\date{\today}

\maketitle

\begin{abstract}
    We propose a topological inflation model in supergravity. In this
    model, the vacuum expectation value (VEV) of a scalar field takes
    a value much larger than the gravitational scale $M_{G} \simeq 2.4
    \times 10^{18}$~GeV, which is large enough to cause topological
    inflation. On the other hand, expansions of the K\"ahler potential
    and the superpotential beyond the gravitational scale are
    validated by the introduction of a Nambu-Goldstone-like shift
    symmetry. Thus, topological inflation inevitably takes place in
    our model.
\end{abstract}

\pacs{PACS numbers: 98.80.Cq,04.65.+e,11.27.+d}


\section{Introduction}

Inflation is the most powerful extension to the standard big bang
theory because it gives solutions to the flatness problem, the horizon
problem, the origin of density fluctuations, and so on
\cite{inflation}. Recent observation of anisotropies of the cosmic
microwave background radiation (CMB) by the Boomerang
\cite{BOOMERANG}, the MAXIMA \cite{MAXIMA}, and the DASI \cite{DASI}
experiments found the first acoustic peak with a spherical harmonic
multipole $l \sim 200$, which implies the standard inflationary
scenario. Up to now, many types of inflation models have been
proposed. Among them, chaotic inflation \cite{chaotic} is very
attractive because it does not suffer from any initial condition
problem, especially, the flatness (longevity) problem. All other
models which occur at low energy scales sustain this problem.  That
is, why does the universe live so long up to the low energy scale from
the gravitational scale? However, this problem is evaded if the
universe is open at the beginning. While new inflation and hybrid
inflation have another severe problem, namely, the initial value
problem \cite{inflation,inithybrid}, a fine-tuning of the initial
value of the inflaton is not needed for topological inflation
\cite{Linde,Vilenkin}. Thus, topological inflation is still attractive
if the universe is open at the beginning. In fact, a possibility is
pointed out that the quantum creation of an open universe can take
place with appropriate continuation from the Euclidean instanton
\cite{open}. Furthermore, the spectrum of density fluctuations
predicted by topological inflation becomes a tilted one, which may be
testable in galaxy surveys and CMB observations.

Supersymmetry (SUSY) is one of the most powerful extensions to the
standard model of particle physics because it stabilizes the
electroweak scale against radiative corrections and realizes the
unification of the standard gauge couplings. Therefore, it is
important to consider inflation models in the context of SUSY and its
local version, i.e., supergravity (SUGRA). In the context of SUGRA,
topological inflation is again favorable because it straightforwardly
predicts the reheating temperature low enough to avoid the
overproduction of gravitinos for a wide range of the gravitino mass.
This is mainly because topological inflation occurs at a low energy
scale and the inflaton has only gravitationally suppressed
interactions with standard particles to keep the flatness of the
potential. Furthermore, it is attractive in the scheme of superstring
theories\footnote{The superstring inspired models of topological
inflation were studied in \cite{BBN,EKOY}.} because superstring
theories compactified on $(3+1)$-dimensional space-time have many
discrete symmetries in the low-energy effective Lagrangian
\cite{discrete}, which is very useful to cause topological inflation.

Izawa, Kawasaki, and Yanagida proposed a topological inflation model
in supergravity with an $R$-invariant vacuum, which guarantees the
vanishing cosmological constant at the end of inflation \cite{IKY}.
However, the model has two weak points, which come from the same
source. First of all, the K\"ahler potential is expanded around the
origin with expansion parameter $|\phi|/M_{G}$. However, the critical
value $\phi_{c}$ of the vacuum expectation value (VEV) $\la \phi \ra$
of the inflaton to cause topological inflation is roughly the
gravitational scale $M_{G}$. This can be understood from a simple
discussion \cite{Linde,Vilenkin}. The typical radius $r \sim \la \phi
\ra/v^{2}$ of the topological defect is given by equating the gradient
energy $(\la \phi \ra/r)^{2}$ and the potential energy $v^{4}$.  For
topological inflation to occur, the typical radius $r$ must be larger
than the hubble radius given by $H^{-1} \sim M_{G}/v^{2}$, which leads
to the rough condition $\la \phi \ra \gtrsim M_{G}$. In fact, Sakai
{\it et al}. found the critical value $\phi_{c} \simeq 1.7 M_{G}$
irrespective of the coupling constant for a double well potential
\cite{sakai}. Later, the supergravity model \cite{IKY} was
investigated in detail and it was found that the critical value
$\phi_{c}$ slightly depends on the slope of the potential and is as
small as $0.95 M_{G}$ at best \cite{KSYY}. Thus, one wonders if the
expansion of the K\"ahler potential is valid.  Of course, the
$R$-invariant vacuum given by the requirement of $\del W / \del
\phi_{i} = W = 0$ for all scalar fields $\phi_{i}$ is unchanged
irrespective of the form of the K\"ahler potential. So, the
description is still valid that the potential is flat around the
origin and the global minima are given by the $R$-invariant vacua.
However, as the inflaton approaches the gravitational scale, the
expansion of the K\"ahler potential becomes invalid. Thus, the
potential may take a nontrivial shape beyond the gravitational scale
and, even worse, a barrier may appear between the flat slope around
the origin and the global minima so that inflation becomes the type of
old inflation, which results in an inhomogeneous universe and hence
does not work as an inflation model.

A similar problem may apply to the superpotential. In Ref.
\cite{IKY}, the superpotential is truncated up to the quadratic term
($\lambda'\phi^{2}$ : $\lambda'$ a real constant) of the inflaton
superfield for simplicity. However, higher order terms
($\CO[(\lambda'\phi^{2})^{n}]$ or $\CO[(\phi^{2})^{n}]$, $n \ge 2$)
may appear, which may drastically change the shape of the potential
around the global minima again. Thus, the model proposed in Ref.
\cite{IKY} may not work.

In this paper, we propose a new model of topological inflation in
supergravity, where the above problems are evaded by the introduction
of a Nambu-Goldstone-like shift symmetry \cite{KYY}. In the next
section, we give our model of topological inflation. For successful
topological inflation, we introduce symmetries and, if necessary,
spurion fields whose vacuum values softly break the introduced
symmetries. In Sec. III, we investigate the dynamics of topological
inflation in detail and give a constraint on the parameters associated
with the vacuum values of the spurion fields. In the final section, we
give the summary of our results.

\section{Model of topological inflation}

First of all, we introduce a $Z_{2}$ symmetry as an example of a
discrete symmetry, which is necessary for producing domain walls.  We
assume that the inflaton supermultiplet $\Phi$ is odd and the other
supermultiplets introduced later are even under the above $Z_{2}$
symmetry. As the inflaton field $\varphi$ acquires its vacuum
expectation value (VEV), the $Z_{2}$ symmetry is spontaneously broken.

Next, we introduce a Nambu-Goldstone-like shift symmetry to validate
the expansion of the K\"ahler potential. We assume that the model is
invariant under the following Nambu-Goldstone-like shift symmetry
\cite{KYY}: $\Phi \rightarrow \Phi + C M_{G}$, where $C$ is a
dimensionless real constant.\footnote{A shift symmetry introduced by
us is possessed by so-called no-scale supergravity theories, which
appear in the low energy limit of superstring theories. However,
inflation does not take place for a simple no-scale type K\"ahler
potential. So, in order to realize inflation, we will slightly change
the form of the K\"ahler potential keeping the shift symmetry. We hope
that our inflaton is one of modulus fields in string theories after we
reveal dynamics of the string, particularly, the compactification
mechanism, which has not been known yet.} Then, the K\"ahler potential
is a function of $\Phi - \Phi^{\ast}$, i.e., $K(\Phi,\Phi^{\ast}) =
K(\Phi - \Phi^{\ast})$, which allows the real part of the scalar
components of $\Phi$ to take a value larger than the gravitational
scale. However, if the shift symmetry is exact, the inflaton cannot
have any potential. So, we need to break it softly for successful
inflation.  For this purpose, we introduce a spurion field $\Xi$ and
extend the shift symmetry to include the spurion field $\Xi$. We
assume that the model is invariant under
\bea
  \Phi &\rightarrow& \Phi + C M_{G}, \non \\
  \Xi  &\rightarrow& \lmk\frac{\Phi}{\Phi + C M_{G}}\rmk^{2} \Xi.
  \label{eq:shift}
\eea
That is, the combination $\Xi\Phi^{2}$ is invariant under the shift
symmetry. The vacuum value of $\Xi$, $\la \Xi \ra = u \ll 1$, softly
breaks the shift symmetry. Here and hereafter, we set $M_{G}$ to be
unity.

Furthermore, we introduce the $U(1)_{\rm R}$ symmetry ($R$ symmetry)
because it prohibits a constant term in the superpotential, which
ensures vanishing cosmological constant at the end of inflation. Since
the K\"ahler potential is invariant only if the $R$ charge of $\Phi$ is
zero, another supermultiplet $X(x,\theta)$ with its $R$ charge equal to
2 must be introduced. Then, the superpotential invariant under the
$Z_{2}$, the shift and the $U(1)_{\rm R}$ symmetries is given by
\beq
  W = X [\,\alpha_{0} + \alpha_{1} \Xi\Phi^{2} + 
           \alpha_{2} (\Xi\Phi^{2})^{2} + \cdots\,]
  \label{eq:superpotential}
\eeq
with $\alpha_{i} = \CO(1)$. As shown later, for successful topological
inflation, the coefficient of $X$, $\alpha_{0}$, must be suppressed.
For this purpose, we introduce another $Z_{2}$ symmetry (named
$Z'_{2}$) and another spurion field $\Pi$. Under this $Z'_{2}$
symmetry, $\Phi$ is even and the other superfields are odd. The vacuum
value of $\Pi$, $\la \Pi \ra \equiv v \ll 1$, softly breaks the
$Z'_{2}$ symmetry so that the smallness of $\alpha_{0} = v$ is
associated with the breaking of the $Z'_{2}$ symmetry. Here you should
notice that the $Z'_{2}$ charge of the spurion field $\Xi$ is also
odd. Then, two cases are possible. In the first case, the $Z_{2}$ and
the shift symmetries are broken at the same time. In this case, we
expect $\CO(u) = \CO(v)$. A similar case is discussed in Ref.
\cite{Yamaguchi} in the context of double inflation. In the second
case, the $Z'_{2}$ symmetry is broken first and later the shift
symmetry is broken. In this case, we expect $\CO(u) \ll \CO(v)$.  In
this paper, we assume the second case. Then, inserting the vacuum
values of the spurion fields, the superpotential is written as
\beq
  W = X [\,v - u \Phi^{2} +  
           \alpha_{2} (u \Phi^{2})^{2} + \cdots\,].
  \label{eq:superpotential2}
\eeq
The higher order terms $\alpha_{i} (u \Phi^{2})^{i} : i \ge 2$ are
negligible because we are interested in only the field values up to
the VEV of $\Phi$, that is, $u |\Phi^{2}| \sim v \ll 1$. After all, we
take the following superpotential:
\beq
  W = v X ( 1 - g \Phi^{2} ),
  \label{eq:superpotential3}
\eeq
with $g \equiv u / v \ll 1$. Here we have assumed that both constants
$u$ and $v$ are real and positive for simplicity.

In the same way, the K\"ahler potential neglecting higher order terms
is given by
\beq
 K = - \frac12\,(\Phi - \Phi^{\ast})^{2} + |X|^{2}.
  \label{eq:Kahler}
\eeq
The higher order terms such as $u'(\Phi^{2} + \Phi^{\ast\,2})$ with
$\CO(u) = \CO(u')$ are negligible because $u \ll 1$ and we are
interested in only the field values up to the VEV of $\Phi$, that is,
$u |\Phi^{2}| \sim v \ll 1$. The charges of the supermultiplets are
shown in Table \ref{tab:charges}. Here it should be noticed that the
model is natural in the sense of 't Hooft \cite{tHooft}, that is, the
symmetries are recovered if the small parameters $u$ and $v$ are set
to be zero.

\section{Dynamics of topological inflation}

In this section we investigate the dynamics of topological inflation
and give a constraint on the parameters associated with the breaking
of the symmetries.

The superpotential and the K\"ahler potential given in the previous
section lead to the Lagrangian density $\CL(\Phi,X)$ for the scalar
fields $\Phi$ and $X$,
\beq
  \CL(\Phi,X) = \partial_{\mu}\Phi\partial^{\mu}\Phi^{\ast} 
       + \partial_{\mu}X\partial^{\mu}X^{\ast}
         -V(\Phi,X),
  \label{eq:lagrangian}
\eeq
where the scalar potential $V$ is given by
\beq
  V = v^{2} e^{K} \lkk\,
      \left|\,1 - g\Phi^{2}\,\right|^{2}(1-|X|^{2}+|X|^{4}) 
       + |X|^{2} \left
         |\,2g\Phi + (\Phi-\Phi^{\ast})(1-g\Phi^{2})\,
                 \right|^{2}
                  \,\rkk.
    \label{eq:potential}
\eeq
Here and hereafter, we denote the scalar components of the
supermultiplets by the same symbols as the corresponding
supermultiplets.

We decompose the scalar field $\Phi$ into the real component $\varphi$
and the imaginary component $\chi$,
\beq
  \Phi = \frac{1}{\sqrt{2}}\,(\varphi + i \chi).
  \label{eq:decomposition}
\eeq
Then the Lagrangian density $\CL(\varphi,\chi,X)$ is written as
\beq
  \CL(\varphi,\chi,X) = 
              \frac{1}{2}\,\partial_{\mu}\varphi\partial^{\mu}\varphi 
              + \frac{1}{2}\,\partial_{\mu}\chi\partial^{\mu}\chi 
              + \partial_{\mu}X\partial^{\mu}X^{*}
              -V(\varphi,\chi,X)
  \label{eq:lagrangian2}
\eeq
with the potential $V(\varphi,\chi,X)$ given by
\bea
  V(\varphi,\chi,X)
    &=& v^{2}
           \exp \lkk \chi^{2} + |X|^{2}\,
                \rkk \non \\ 
    && \hspace{-2.0cm} \times
         \lkk\,\lhk\,\lmk 
              1 - \frac{g}{2}\,\varphi^{2} 
                     \rmk^{2} 
                + \chi^{2} \lkk g + 
                     \frac{g^{2}}{4}\,(2 \varphi^{2} + \chi^{2})
                           \rkk\,
                \rhk\,
                  (1-|X|^{2}+|X|^{4}) 
         \right. \non \\ 
    && \hspace{-1.5cm}
             +~|X|^{2} 
              \lhk\,
                2\,g^{2}\,(\varphi^{2}+\chi^{2})
              \right. \non \\
    && \hspace{-0.0cm} \left.  \left.
                + 4 g \chi^{2} \lkk 
                       1 + \frac{g}{2}\,(\varphi^{2}+\chi^{2})
                                \rkk
                + 2 \chi^{2} \lkk 1 - g\,(\varphi^{2}-\chi^{2})           
                 + \frac{g^{2}}{4}\,(\varphi^{2}+\chi^{2})^{2}
                             \rkk
              \rhk\,
          \rkk.
  \label{eq:potential2}
\eea

Due to the exponential factor $e^{\chi^{2}+|X|^{2}}$, $\chi$ and $|X|$
are at most of the order of unity. On the other hand, $\varphi$ can
take a value much larger than unity without costing exponentially
large potential energy. Then, $\varphi$ may take a value $\varphi =
\CO(1/\sqrt{gv}) = \CO(1/\sqrt{u})$ in some regions of the universe,
another value $\varphi = \CO(1/\sqrt{g})$ in other regions, and so on.
The cosmic history is quite different according to the initial value
of $\varphi$. For example, in the region with the value $\varphi =
\CO(1/\sqrt{u})$, the term $\frac{u^{2}}{4}\varphi^{4}$ dominates the
scalar potential, which causes chaotic inflation. Furthermore, if we
fine-tune the initial value of $\varphi$, it passes through the global
minimum $\la \varphi \ra = \pm \sqrt{2/g}$ after chaotic inflation and
stops near the local maximum $\varphi = 0$ so that new inflation may
take place. Because of the peculiar nature of new inflation,
primordial black holes may be produced. Thus, chaotic-new inflation
may take place. However, the case with the similar potential has been
already discussed in Ref. \cite{chaonew}. So we concentrate on another
interesting region in this paper.

In other regions, there are some places with the values $\varphi
\simeq \sqrt{2/g}$ and $\varphi \simeq - \sqrt{2/g}$. Then,
topological defects (domain walls) may form if the energy density of
the universe dropped enough. In fact, if the VEV of $\varphi$ is
larger than the gravitational scale $M_{G}$ ($g \lesssim 1$), the
topological defect is unstable and the universe expands exponentially,
that is, topological inflation takes place \cite{Linde,Vilenkin}. In
this paper, we investigate the dynamics of this topological inflation
in detail. Here, one should notice that in other regions chaotic
inflation takes place just below the Planck scale and the universe
expands enough so that our topological inflation model can be free
from the flatness problem too. The observations such as spectral index
of density fluctuations and gravitational waves decide which region
the present universe belongs to.

The effective mass squared of $\chi$, $m_{\chi}^{2}$ during
topological inflation is given by
\beq
  m_{\chi}^{2 }\simeq 6 H^{2},
  \label{eq:masssquared}
\eeq
where $H$ is the hubble parameter given by $H^{2} \simeq v^{2}/3$.
Thus, once topological inflation takes place, $\chi$ rapidly
oscillates around the origin and the amplitude decays in proportion to
$a^{-3/2}$ ($a$ : the scale factor). Therefore, we can safely set
$\chi$ to be zero at least classically.

Using $\chi \ll 1$, the scalar potential is approximated as
\bea
  V &\simeq& v^{2} \lkk\,\lmk 1 - \frac{g}{2} \varphi^{2} \rmk^{2} 
                    + 2 g^{2} \varphi^{2} |X|^{2}\,\rkk \non \\
    &\simeq& v^{2} \lmk\,1 - g \varphi^{2} 
               + 2 g^{2} \varphi^{2} |X|^{2}\,\rmk
             \qquad \qquad \qquad {\rm for} \quad \varphi \ll 1.
  \label{eq:potential3}
\eea
The effective mass squared of $X$, $m_{X}^{2}$, is given by
\bea
  m_{X}^{2} &\simeq& 2 g^{2} v^{2} \varphi^{2} \non \\
            &\simeq& 6 g^{2} \varphi^{2} H^{2} \ll H^{2}.
  \label{eq:masssquared2}
\eea
Thus, $X$ does not oscillate around the origin and instead slow-rolls
down along the potential. In fact, during topological inflation,
\beq
  X \sim X_{i} \exp \lmk - \frac{g}{2} \varphi^{2} \rmk,
  \label{eq:Xvalue}
\eeq
where $X_{i}< 1$ is the value of $X$ at the beginning of topological
inflation. Thus, $|X| < 1$ throughout topological inflation and $|X|
\ll 1$ near the end of topological inflation.\footnote{If we take into
account a higher order term $ - \frac{k_{1}}{4}|X|^{2}$ ($k_{1}
\gtrsim 1$) in the K\"ahler potential, the effective mass squared of
$X$ is much larger than $H^{2}$ so that the amplitude of $X$ rapidly
decays and throughout topological inflation we can safely set $X$ to
be zero at least classically.} The last term in the potential
(\ref{eq:potential3}) is irrelevant for the dynamics of $\varphi$
because $g\ll 1$ and $|X| < 1$.

Using the slow-roll approximation, the $e$-fold number acquired for
$\varphi > \varphi_N$ is given by
\beq
  N \simeq \int_{\varphi_{f}}^{\varphi_{N}} \frac{V}{V'} 
    \simeq \frac{1}{ 2g } 
              \ln \lmk \frac{\varphi_{f}}{\varphi_{N}} \rmk, 
  \label{eq:efold}
\eeq
where the prime represents the derivative with respect to $\varphi$
and $\varphi_{f} \sim \sqrt{2/g}$ is the value of $\varphi$ at the end
of topological inflation. $\varphi_N$ is represented by the $e$-fold $N$ 
as
\beq
  \varphi_{N} \sim \varphi_{f} e^{-2 g N} 
               \sim \sqrt{\frac{2}{g}}\,e^{-2 g N}.
  \label{eq:varphiN}
\eeq

Next we evaluate the density fluctuations produced during topological
inflation. In this model, there are two effectively massless fields
$\varphi$ and $X$ during topological inflation. However, we can easily 
show that the metric perturbation in the longitudinal gauge
$\Phi_A$ can be estimated as \cite{PS}
\bea
  \Phi_A &=& - \frac{\dot{H}}{H^{2}} C_{1}
               - 2 g^{2}{\varphi^{2}}X^{2} C_{3}, \non \\
  C_{1} &=& H \frac{\delta\varphi}{\dot{\varphi}}, \non \\
  C_{3} &=& H \lmk \frac{\delta\varphi}{\dot{\varphi}} 
                  - \frac{\delta X}{\dot{X}}
              \rmk 
               g \varphi^{2},
\eea
where the dot represents the time derivative, the term proportional to
$C_{1}$ corresponds to the growing adiabatic mode, and the term
proportional to $C_{3}$ the nondecaying isocurvature mode. For
simplicity, we deal with $X$ as if it is a real scalar field.  You
should notice that only $\varphi$ contributes to the growing adiabatic
fluctuations. Then, the amplitude of the curvature perturbation
$\Phi_A$ on the comoving horizon scale at $\varphi=\varphi_{N}$ is
estimated by the standard one-field formula as
\beq
  \Phi_A \simeq  \frac{f}{2\sqrt{3}\pi} \frac{V^{3/2}}{V'}
            \simeq  \frac{f}{2\sqrt{3}\pi}
                       \frac{v}{2g\varphi_{N}}
\eeq
with $f=3/5~(2/3)$ in the matter (radiation) domination. From the
Cosmic Background Explorer (COBE) normalization $\Phi_A \simeq 3\times
10^{-5}$ at $N\simeq 60$ \cite{COBE}, the vacuum value $v$ is
constrained as
\beq
  v \simeq  1.1 \times 10^{-4} \sqrt{ g } 
    e^{-gN/2}|_{N=60}
    \simeq 3.3 \times 10^{-5} - 6.1 \times 10^{-8}
\eeq
for $0.01 \le g \le 0.05$. The spectral index $n_s$ is given by
\beq
    n_s \simeq 1 - 4 g.
\eeq
Since the COBE data also show $n_s = 1.0 \pm 0.2$ \cite{COBE}, the
parameter $g = u / v$ is constrained as $g \le 0.05$.\footnote{In Ref.
\cite{IKY}, the combination of the parameters must be fine-tuned to
be nearly equal to unity in order to satisfy the constraint from the
spectral index. On the other hand, in our model, the parameter $g$ has
only to be small, which originates from the difference of the breaking
scales of the shift and the $Z'_{2}$ symmetries.}

After topological inflation ends, the inflaton rapidly oscillates
around the global minimum $\la \varphi \ra \equiv \pm \sqrt{2/g}$ and
decays into standard particles, which reheats the universe. The decays
of the inflaton into standard particles can take place if we introduce
the following superpotential,\footnote{The inflaton may also decay
into standard particles if we consider higher order terms
$u''(\Phi^{2} + \Phi^{\ast\,2}) |\psi_{i}|^{2}$ in the K\"ahler
potential. Here $u''$ is a constant associated with the breaking of
the shift symmetry with $\CO(u) = \CO(u'')$ and $\psi_{i}$ are the
standard particles. Then, the reheating temperature $T_{R}$ becomes
$100$ GeV $-$ $50$ MeV for $0.01 \le g \le 0.05$.}
\beq
  W = v' X H_{u}H_{d},
\eeq
where $v'= \alpha \la \Pi \ra$ with $\alpha = \CO(1)$ [$\CO(v') =
\CO(v)$] is a constant associated with the breaking of the $Z'_{2}$
symmetry and $H_{u}, H_{d}$ are a pair of the Higgs doublets. If we
set the $R$ charge and the $Z_{2}$ and the $Z'_{2}$ charges of
$H_{u}H_{d}$ to be zero and even, the above superpotential is
invariant under all the introduced symmetries before inserting the
vacuum value of the spurion field $\Pi$. Then the inflaton $\varphi$
has the interaction with a pair of the Higgs doublets,
\beq
  \CL_{int} \simeq g v v' \la \varphi \ra \varphi H_{u} H_{d},
\eeq
which gives the decay rate
\beq
  \Gamma \sim g^{2} v^{2} v'^{2} \la \varphi \ra^{2} / m_{\varphi}
\eeq
with the mass of the $\varphi$ field, $m_{\varphi} \simeq 2 \sqrt{g}
v$. Then the reheating temperature $T_{R}$ is given by
\bea
  T_{R} &\sim& 0.1 g v v' \la \varphi \ra / \sqrt{m_{\varphi}} \non \\
        &\sim& 5 \times 10^{8}\,{\rm GeV} - 2 \times 10^{6}\,{\rm GeV}
\eea
for $0.01 \le g \le 0.05$. The above reheating temperature is low
enough to avoid overproduction of gravitinos for a wide range of the
gravitino mass \cite{Ellis,MMY}.

\section{Summary}

In the present paper, we have proposed a topological inflation model
in supergravity. In the model, where a discrete symmetry is
spontaneously broken, the vacuum expectation value (VEV) of the scalar
field takes a value much larger than the gravitational scale so that
topological inflation can take place. Generally speaking, expansions
of the K\"ahler potential and the superpotential around the origin are
invalid beyond the gravitational scale. In our topological inflation
model, the expansion of the K\"ahler potential is validated by the
introduction of a Nambu-Goldstone-like shift symmetry. Furthermore,
one can make the expansion of superpotential valid beyond the
gravitational scale by introducing a $Z'_{2}$ symmetry and combining
it with the shift symmetry. Thus, topological inflation inevitably
takes place in our model. Our topological inflation model predicts the
tilted spectrum of density fluctuations, which may be detectable in
future CMB observations and galaxy surveys. Furthermore, the reheating
temperature is low enough to avoid overproduction of gravitinos for a
wide range of the gravitino mass, especially, that predicted in the
gauge mediated SUSY breaking model.

\subsection*{ACKNOWLEDGMENTS}

M.K. is supported in part by the Grant-in-Aid, Priority Area
``Supersymmetry and Unified Theory of Elementary Particles''(No. 707).
M.Y. was partially supported by the Japanese Grant-in-Aid for
Scientific Research from the Ministry of Education, Culture, Sports,
Science, and Technology.

\begin{table}[t]
  \begin{center}
    \begin{tabular}{| c | c | c | c | c | c |}
                   & $\Phi$ & $X$ & $\Xi$ & $\Pi$ & $H_{u}H_{d}$ \\
        \hline
        $Q_R$      & 0      & 2   & 0     & 0     & 0 \\ 
        \hline 
        $Z_{2}$    & $-$    & $+$ & $+$   & $+$   & $+$\\  
        \hline 
        $Z'_{2}$   & $+$    & $-$ & $-$   & $-$   & $+$
    \end{tabular}
    \caption{Charges of the $U(1)_{\rm R} \times Z_{2} \times
    Z'_{2}$ symmetries for the various supermultiplets.}
    \label{tab:charges}
  \end{center}
\end{table}

\end{document}